\documentclass[pra,twocolumn,showpacs,superscriptaddress, nofootinbib]{revtex4-1} 
\usepackage{graphicx,amsmath,bm, color}
\usepackage{latexsym}
\usepackage{amssymb}
\usepackage{amsfonts}
\usepackage{amsthm}
\usepackage{mathrsfs}
\usepackage{natbib}
\usepackage{hyperref}
\usepackage{cleveref}
\usepackage{verbatim,graphics, float}
\usepackage{psfrag}
\usepackage{dsfont}
\usepackage{mathtools}

\newcommand{\ba}{\begin{eqnarray}}
\newcommand{\ea}{\end{eqnarray}}
\newcommand{\ban}{\begin{eqnarray*}}
\newcommand{\ean}{\end{eqnarray*}}

\begin{document}

\title{Optimal Photon Generation from Spontaneous Raman Processes in Cold Atoms}
\date{\today}

\author{Melvyn Ho}
\affiliation{Department of Physics, University of Basel, Klingelbergstrasse 82, 4056 Basel, Switzerland}

\author{Colin Teo}
\affiliation{Department of Physics, University of Basel, Klingelbergstrasse 82, 4056 Basel, Switzerland}
\affiliation{ Centre for Bioimaging Sciences, Department of Biological Sciences, National University of Singapore, 14 Science Drive 4, Singapore 117543}

\author{Hugues de Riedmatten}
\affiliation{ICFO-Institut  de  Ciencies  Fotoniques,  The  Barcelona  Institute  of
Science  and  Technology,  08860  Castelldefels  (Barcelona),  Spain}
\affiliation{ ICREA-Instituci\'o Catalana de Recerca i Estudis Avan\c{c}ats, 08015 Barcelona, Spain}

\author{Nicolas Sangouard}
\affiliation{Department of Physics, University of Basel, Klingelbergstrasse 82, 4056 Basel, Switzerland}

\begin{abstract}
Spontaneous Raman processes in cold atoms have been widely used in the past decade for generating single photons. Here, we present a method to optimize their efficiencies for given atomic coherences and optical depths. We give a simple and complete recipe that can be used in present-day experiments, attaining near-optimal single photon emission while preserving the photon purity. 

\end{abstract}
\maketitle

\section{Introduction }

\begin{figure}[htbp!]
\includegraphics[
width = 1.00 \linewidth]{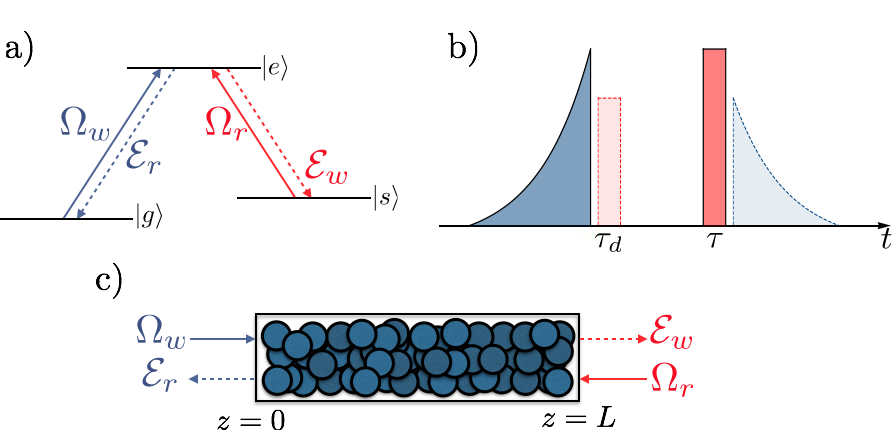}
\caption{Level scheme and schematic of the proposed single photon source. (a) Write (read) control fields are indicated with Rabi frequencies $\Omega_W$ ($\Omega_R$) and write (read) photon fields are indicated with quantum fields $\mathcal{E}_w$ ($\mathcal{E}_r$), each along their respective transitions. In our model the excited level $|e \rangle$ is capable of spontaneous emission to the metastable states $|g\rangle$ and $|s \rangle$. (b) A schematic of the protocol indicates the sequence of events. A fast resonant write control field of duration $\tau_W$ followed by a write photon field detection in a short time window $\tau_d$ heralds a spatially varying spin wave. A fast $\pi$-pulse of duration $\tau_R$ then enables the retrieval of the stored excitation. Laser pulses are shaded darker to indicate their stronger intensities as compared to the weaker photon emissions. (c) Backward retrieval configuration with counterpropagating control fields results in photon field emissions in opposing directions.}\label{quicksumm}
\end{figure}

 On-demand single photon sources are appealing ingredients for many quantum information tasks. Examples include the distribution of entanglement over long distances using quantum repeaters or quantum communications with security guarantees which remain valid, independent of the details of the actual implementation \cite{Eisaman11,Sangouard12}. These tasks necessitate stringent purity and efficiency requirements on the performance of the single photon sources used. Techniques based on spontaneous Raman processes in cold atoms are among the most advanced single-photon sources with such characteristics. The basic principle is to use an ensemble of three-level atoms in a $\Lambda$-configuration and two pulsed laser fields (see Fig \ref{quicksumm}a). The first write pulse -- the write control field -- off-resonantly excites one transition, which can spontaneously produce a frequency-shifted photon -- the write photon field -- along the second transition through a Raman process. Since all the interacting atoms participate in the process, and there is no information about which atom emitted the photon, the detection of this write photon heralds the existence of a single delocalised excitation across the sample -- an atomic spin wave. Once the spin wave has been prepared, the atomic sample is ready to be used as a source, and a second pulse -- the read control field -- along the second transition performs a conversion of the atomic spin wave into a second photon -- the read photon field. If the duration of the process is short enough with respect to the atomic coherence times, and the optical depth of the sample sufficiently high, then the read photon is emitted efficiently in a well defined mode and the protocol provides a viable single photon source.

Such sources have been at the core of numerous experiments during the last decade following the seminal paper of Duan, Lukin, Cirac and Zoller \cite{DLCZ01}, showing how they could be used for long-distance quantum communication based on quantum repeater architectures (for reviews, see \cite{Sangouard11, Simon10, Bussieres13, Heshami16}).  Recently, they have been used as quantum memories with storage times up to 200ms \cite{Radnaev10, Yang16} or as a source producing pure single photons with a temporal duration that can be varied over up to 3 orders of magnitude while maintaining constant efficiencies \cite{Farrera16}. We stress that the efficiency of such a source is a critical parameter for the implementation of efficient quantum repeater architectures. While very high efficiencies of $\sim 90\%$ are essential, a reduction of the source efficiency by 1\% can reduce the repeater distribution rate by 10-20\%,  depending on the specific architecture \cite{Sangouard11}.

Several solutions can be envisioned to ensure high efficiencies. One solution relies on the use of an optical cavity to enhance the spinwave--light conversion efficiency. Experimental efforts along this direction have resulted in efficiencies of up to 84\% \cite{Simon07, Bimbard14}. An alternative solution involves increasing the atomic density in order to obtain a larger optical depth. This however makes operations like optical pumping and noise free operations more challenging. This naturally raises the following question: \emph{What is the optimal efficiency that can be achieved with a bulk atomic ensemble having a certain optical depth?} This question has been previously addressed for memory protocols where single photons are first absorbed before subsequently retrieved in a well defined mode \cite{GorshkovPRL, GorshkovII}. Inspired by these works, we first examine the conditions on the spin wave shape for achieving optimal photon retrieval efficiencies given the optical depth and specified energy levels in the atomic species. After finding the optimal spin wave shapes, we recognise that current approaches using off-resonant write control fields create non-ideal flat spin excitations in the sample (previously studied in works such as \cite{Mendes13}), since such control fields do not experience significant intensity depletion during propagation. To achieve better retrieval efficiencies, we then propose (see Fig \ref{quicksumm}b) to spatially shape the spin wave using resonant, temporally shaped write control fields. Combined with fast read control fields during the retrieval process, we show that our recipe achieves near-optimal retrieval efficiencies and perfect purity.

This paper is structured as follows: In the first section we discuss the optimal retrieval efficiency from a spin excitation. For completeness, we first quickly review derivations in \cite{GorshkovII} that allow us to find the expression for the retrieval efficiency of a complete retrieval process, where we begin with only $|g\rangle $-$|s\rangle $ coherences and transfer all atoms to $|g\rangle $. We then find the shapes of the spin excitation that yield the optimal retrieval efficiency when complete retrieval is performed. In the second section we propose the use of a resonant write control field to create spin excitations similar to those that allow for optimal retrieval. We then give explicit expressions for retrieval when using a quick read control field with a constant Rabi frequency. Finally, we include a feasibility study in the case of a gas of Rubidium-87.

\section{optimal retrieval}
\subsection{Efficiency of a complete retrieval process}

To begin our analysis, we first review a derivation in \cite{GorshkovII} giving the efficiency of the retrieval process on the spin shapes of the atomic excitation. We emphasise that the work in \cite{GorshkovII} focuses on absorptive memory protocols where a field is first absorbed in an atomic medium, creating a spin wave that can be read out later to re-emit the field in a well defined spatio-temporal mode. 
In our proposal, the spin wave creation is instead heralded by the detection of the write photon field, but the readout process is analogous, allowing us to make use of Ref \cite{GorshkovII} to deduce the spin wave shapes that maximise the retrieval efficiency.

We consider a three-level atomic system in a $\Lambda$-configuration (see Fig \ref{quicksumm}a) with spin excitations present in the form of $|g\rangle $-$|s\rangle $ coherences. In the situation where almost all the atoms remain in $|g\rangle $ and in a rotating frame, the backward wave propagation equation (see Fig 1c) along with the Heisenberg-Langevin equations of motion yield 
\begin{align}
\partial_z \mathcal{E}_r(z,t) =&  -i \sqrt{\frac{d \gamma_{eg}}{c L}} P(z,t), \nonumber \\
\partial_t P(z,t) =& -(\gamma_{eg} + i \Delta)P(z,t) + i \sqrt{\frac{d \gamma_{eg} c}{L}} \mathcal{E}_r(z,t) \nonumber  \\
& + i \Omega_R (t) S(z,t) + F_P(z,t), \nonumber \\
\partial_t S(z,t) =& - \gamma_0 S(z,t)+  i \Omega_R ^*(t) P(z,t)  + F_S(z,t), \label{HeisenbergLangevinRetrieval}
\end{align}

\noindent  where $P(z,t) = \sqrt{N} \sigma_{ge} (z,t) e^{-i \omega_1 \frac{L-z}{c}}$ and $S(z,t) = \sqrt{N} \sigma_{s} (z,t) e^{-i (\omega_1-\omega_{2}) \frac{L-z}{c}}$  are rescaled and slowly varying atomic operators (see Appendix for details), with $\omega_{1}$ ($\omega_{2}$) referring to the energy transition of the $|e\rangle $-$|g\rangle $ ($|e\rangle $-$|s\rangle $) transition. $\gamma_{eg}$ ($\gamma_0$) refers to the decay rate of the $|e\rangle $-$|g\rangle $ ($|g\rangle $-$|s\rangle $) transition. $L$ denotes the length of the atomic sample and $N$ the number of atoms within this sample. $F_S$ and $F_P$ indicate the noise operators associated to $S$ and $P$ respectively.
$\Omega_R$ ($\Delta$)  refers to the Rabi frequency (detuning) of the classical write control field on the $|e\rangle $-$|g\rangle $ transition, and $\mathcal{E}_r$ denotes the quantum field of the retrieval emission. The optical depth $d$ characterises the absorption of resonant light in the sample, such that the outgoing light intensity is $I_0(z=L) = e^{-2d}I(z=0)$,  valid when the spectrum of the incoming light is well contained within the atomic bandwidth.

Here, we consider the situation where retrieval is completed well within the spin wave decoherence time, and thus ignore $\gamma_0$. We also ignore the noise terms $F_S$ and $F_P$ since they do not contribute to the spin and photon numbers, which are the relevant quantities here.

Defining first the reversed functions  $\bar{P}(L-z,t) = P(z ,t)$, $\bar{S}(L-z,t) = S(z ,t)$ and $\bar{\mathcal{E}}_r(L-z,t) = \mathcal{E}_r(z ,t)$, then taking the Laplace transforms of Eqns (\ref{HeisenbergLangevinRetrieval}) from $L-z =z' \rightarrow u$, we begin with the following set of transformed equations
\begin{align}
\bar{\mathcal{E}}_r(u,t) =& i \sqrt{\frac{ \gamma_{eg} d }{c L } } \frac{1}{u} \bar{P}(u,t),  \label{Etransformeqn}\\
\partial_t \bar{P}(u,t) =& - \Big[ \gamma_{eg} (1+ \frac{d}{L u }) + i \Delta \Big] \bar{P}(u,t) \nonumber \\ 
& + i\Omega_R(t) \bar{S}(u,t), \label{Ptransformeqn} \\
\partial_t \bar{S}(u,t) =& \  i \Omega_R^*(t) \bar{P}(u,t). \label{Stransformeqn}
\end{align}

\noindent From Eqns  (\ref{Ptransformeqn}) and (\ref{Stransformeqn}) we first obtain the following result
\begin{align}
&\frac{d}{d t}\Big(\langle \bar{P} ^\dagger(u_1, t) \bar{P}(u_2, t) +\bar{S}^\dagger(u_1, t) \bar{S}(u_2, t)  \rangle \Big) \nonumber \\
&= \gamma_{eg}\Big(-2 - \frac{d}{L u_1} - \frac{d}{L u_2}\Big) \langle \bar{P}^\dagger (u_1, t) \bar{P}(u_2, t) \rangle.  \label{intermediate}
\end{align}

\noindent With Eqn (\ref{Etransformeqn}) we can then rewrite the number of emitted photons $\eta$ in terms of $P(u,t)$
\begin{align}
\eta  &= \frac{c}{L} \int_{0} ^\infty dt \  \langle \mathcal{E}^\dagger_{r}(z=0, t)\mathcal{E}_r(z=0, t) \rangle  \nonumber \\
&=   \frac{c}{L}  \mathcal{L}_2^{-1}  \int_{0}^\infty d{t} \  \frac{\gamma_{eg} d }{c L}\frac{1}{ u_1 u_2} \langle \bar{P}^\dagger(u_1, {t})  \bar{P}(u_2,{t})\rangle \  \Big{|} _{\ \ \substack{\mathllap{z'_1}\rightarrow \mathrlap{L} \\ \mathllap{z'_2} \rightarrow \mathrlap{L}}} \ \ \  ,  \nonumber 
\end{align}

\noindent where $\mathcal{L}_2^{-1}$ indicates the instruction to take the Laplace inverses of both $u_1$ and $u_2$ separately. With the use of Eqn (\ref{intermediate}) we can next rewrite $\langle \bar{P} ^\dagger(u_1, t) \bar{P}(u_2, t) \rangle$ as a full derivative and perform the integral to get
\begin{align}
\eta &=\mathcal{L}_2^{-1}\frac{d}{L} \frac{-1}{(u_1 + u_2) d + 2Lu_1 u_2} \nonumber\\
& \Big(  \langle \bar{P}^\dagger(u_1, {t})  \bar{P}(u_2,{t})\rangle  + \langle\bar{S}^\dagger(u
_1, {t})  \bar{S}(u_2,{t})\rangle \Big)  \ \Big{|} _{t = 0}  ^{ \infty} \  \Big{|} _{\ \ \substack{\mathllap{z'_1}\rightarrow \mathrlap{L} \\ \mathllap{z'_2} \rightarrow \mathrlap{L}}}  \nonumber \\
&=  \frac{1}{L^2} \mathcal{L}_2^{-1}   \frac{d L}{d(u_1+u_2)  + 2 L u_1 u_2 } \langle \bar{S}^\dagger(u_1, 0) \bar{S}(u_2,0) \rangle  \  \Big{|} _{\ \ \substack{\mathllap{z'_1}\rightarrow \mathrlap{L} \\ \mathllap{z'_2} \rightarrow \mathrlap{L}}} \nonumber \ \ \  ,
\end{align}

\noindent where the last equality comes from the conditions we assume in a complete retrieval process, i.e. that we begin with only $|g\rangle $-$|s\rangle $ coherences and at the end of the process all atoms are in $|g\rangle $.
By performing the inverse Laplace transforms one sees that for complete retrieval in the backward direction\footnote{In Ref \cite{GorshkovII}, Eqn (\ref{CompleteRetrieval}) is said to describe the \emph{optimal retrieval efficiency} from a given spin wave. For us, we see this retrieval efficiency function as a description of \emph{complete retrieval} in the absence of spin wave decoherence, which is made optimal only when provided with the correct spin excitation.},
\begin{align}
\eta =&\frac{1}{L}\int_0 ^L dz_1 \frac{1}{L} \int_0 ^L dz_2  \ k_r(L-z_1, L-z_2)   \nonumber\\
&\langle S^\dagger(L-z_1, 0) S(L-z_2, 0) \rangle,\label{CompleteRetrieval}
\end{align}

\noindent where $k_r(z_1, z_2) = \frac{d}{2} e^{-d \frac{z_1+z_2}{2 L} }I_0\Big(\frac{d}{L} \sqrt{z_1 z_2}\Big)$ and $I_n(x)$ indicates the modified $n$-th Bessel function of the first kind.
We proceed by considering the situation where there is originally a single spin wave in the sample (such that $\frac{1}{L}\int_0 ^L \, S^\dagger(z,0) S(z,0) \ dz =1$), and thus interpret $\eta$ as the efficiency of the retrieval process. The retrieval efficiency $\eta$ is independent of the details of the read control field used, and is a result of the ratio between desired and undesired modes that are retrieved from the spin wave.

\subsection{Optimal spin shapes for complete retrieval}

Having shown the dependence of the retrieval efficiency on the spin wave shape, we now look for the spin shapes that allow one to maximise the retrieval efficiency in the case of complete retrieval. To do this, we recognise Eqn (\ref{CompleteRetrieval}) as the continuous form of a product of discretised versions of $k_r$ (in the form of a matrix) and $|S\rangle$ (in the form of a vector). 

\begin{center}
\begin{figure}[ht!]
\includegraphics[width=0.97\linewidth]{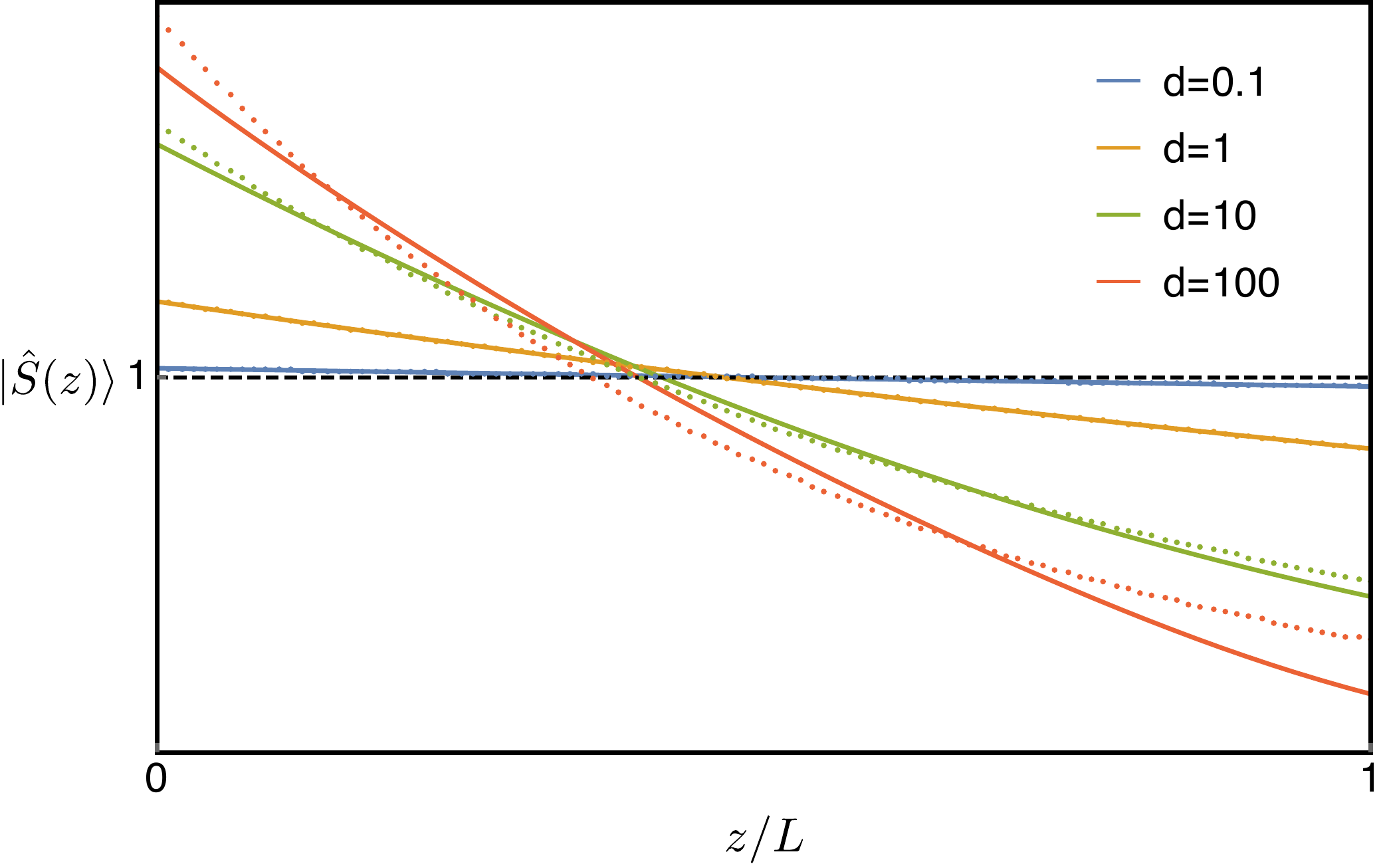}
\caption{Optimised spin wave shapes for retrieval in the backward direction (solid lines) when compared to the best fitting exponential shapes created by our resonant write protocol (dashed lines).}
\label{spinshapes}
\end{figure}
\end{center}

Cast in this light, this integral can be computed by performing a matrix multiplication between the discretised versions of $k_r$ and $|S\rangle $.
In this discrete approximation, the optimal spin shape is thus the eigenvector of $k_r$ with the largest eigenvalue. One can then interpolate the resulting vector to obtain optimised spin shapes, which are presented in Fig \ref{spinshapes}.

The best spin shapes for optimal retrieval show a clear spatial dependence with a bias (depending on the optical depth $d$) towards placing larger excitation probabilities towards the retrieval direction (backwards in this case). We will denote the retrieval efficiencies from these optimal spin shapes as $\eta^*$.

\section{ Practical recipe for achieving near-optimal retrieval efficiencies}
\subsection{Creating spatially varying excitations}

In the previous section, we have outlined how the retrieval efficiency depends on the shape of the given spin excitation, and also how the optimal spin shapes can be computed. Here we propose a method of conveniently creating spin shapes that yield near-optimal retrieval efficiencies. In contrast to creating spin excitations using spontaneous Raman processes enabled by far-detuned write control fields, we explore the use of resonant control fields instead, which create spin excitations with significantly position-dependent excitation profiles.
The ability to send resonant exponential write control fields with different durations further allows us to control the degree of spatial bias that we create in the spin excitation, as we show below.


We give a detailed derivation of the write process in Appendix B1. To summarize (see Fig \ref{quicksumm}b), beginning with all atoms in the $|g\rangle $-level, we send a short rising exponential resonant write pulse with Rabi frequency $\Omega_W(0,t) = \Omega_W ^{max} e^{t/\tau_W} $ that does not significantly excite the atoms to the $|e\rangle $ level ($\Omega_W ^{max} \tau_W \ll 1$). If sent with a sufficiently short duration ($\tau_W \ll 1/\gamma_{eg}$) and shut off at $t=0$, one can consider only the dynamics along the $|g\rangle $-$|e\rangle $ transition, and obtain atomic coherences of the form (see Appendix B2)
\begin{align}
\sigma_{ge} (z,0) =  e^{i k_w . z} \theta_0  e^{- \frac{\alpha z}{2} }
\end{align}
where  $\theta_0 = i \frac{\Omega_W ^{max} \tau_W}{1 + \gamma_{eg} \tau_W}$, $\alpha /2  =  d \frac{\gamma_{eg}\tau_W }{1+ \gamma_{eg}\tau_W}\frac{1}{L}$ and $k_w$ indicates the wave vector for the write photon, which is described using a quantum field $\mathcal{E}_w$.
Immediately after the preparation, we look for the detection of write photons within a short detection window $\tau_d$ as a herald for single spin excitations. This avoids potential dephasing effects from the decoherence of the $|e\rangle $ level. In this short detection window of duration $\tau_d  \ll \text{min}( \frac{1}{2 \gamma_{es}},\frac{1}{2 \gamma_{eg}}$),  and where  $\tau_d \ll \{  {\bar{d} \  \gamma_{es} }  |\theta_0|^2  \frac{1 - e^{-\alpha L}}{\alpha L} \}^{-1}$, ensuring the number of emitted write photons $n_{w}$ is much smaller than 1, we obtain (see Appendix B3, B4)
\begin{align}
n_{w} = 
\big( {\bar{d} \  \gamma_{es} \tau_d} \big )  |\theta_0|^2  \frac{1 - e^{-\alpha L}}{\alpha L} ,
\end{align}
 where $\gamma_{es}$ ($\bar{d}$) refers to the decay rate (optical depth) of the $|e\rangle $-$|s\rangle $ transition. The write photon number $\eta_w$ is simply the product of ${\bar{d} \  \gamma_{es} \tau_d}$ and the fraction of excited atoms (averaged across the sample).
 
In this same regime for $\tau_d$, to leading order the corresponding spin state is (see Appendix B5)
\begin{align}
S ^\dagger(z,\tau_d) =& -i  \sqrt{\frac{\bar{d} \gamma_{es} c}{L}} \theta_0e^{-\alpha z/2} \nonumber \\
&  \int_0^{\tau_d} \! e^{-\gamma_0(\tau_d - t_a)}   \mathcal{E}_w(0,t_a)  \, dt_a , \label{SpinPrep}
\end{align}
\noindent which has an exponentially decaying spatial dependence from the $z=0$ side of the sample. The extent of this spatial decay is characterized by $\alpha$, which does depend on the given properties of the atomic sample, but can be controlled by varying the write control field duration $\tau_W$.

\subsection{Performing fast retrieval}

We now proceed with the retrieval process, and spell out the exact requirements for a certain implementation of retrieval -- the fast $\pi$-pulse using a square waveform of duration $\tau_R$.  Once again, we focus on retrieval processes completed well within the spin wave decoherence time and performed under relevant experimental conditions. We thus ignore both the spin decoherence and Langevin noise terms in Eqn (\ref{HeisenbergLangevinRetrieval}). Here we have implicitly assumed that the energy levels of the $|g\rangle$ and $|s\rangle$ levels are degenerate.\footnote{The phase-matching condition in one dimension is fully satisfied for co-propagating pulses and emissions, even in the non-degenerate case. For counter-propagating strategies like the one we suggest, one requires the condition $|\Delta k| L \ll1$, where $\Delta k  = k_W - k_w (= k_R - k_r)$ refers to the difference in wave vector along our 1-dimensional system for the write (read) control and photon fields (see Appendix C).} See \cite{GorshkovII, Hammerer10} for details.

With a resonant square retrieve pulse in the backward direction (See Fig 1c) one finds the following simple expression for the dynamics of the spin wave (details given in Appendix A1)
\ba
\ddot{\bar{S}} (u,t)  + A \dot{\bar{S}} (u,t) + B \bar{S} (u,t)=0,
\ea

\noindent where $A =\gamma_{eg} (1 + \frac{d}{L u })  $ and $B = \Omega_R^2$ (for real $\Omega_R$), and we have taken the Laplace transform $ L-z=z'\rightarrow u$.

In the regime\footnote{In considering the lossless preparation of $\bar{P}(u,t)$ from $\bar{S}(u,t=\tau_d)$, requiring $2\Omega_R \gg \gamma_{eg}(1+d)$ for the $\pi$-pulse can be demanding. However, we show in Appendix A3 that one can achieve the same retrieval efficiency even in the slow readout regime where we do not separate the P preparation process from the emission.
} where $2\Omega_R \gg \gamma_{eg}(1+d)$, we find $4B \gg A^2$, and obtain the following solution
\ba
\bar{S} (u,t) =  e^{-At/2} \cos(\Omega_R t) \bar{S}(u,t=\tau_d),
\ea
\noindent which yields the following expression
\begin{align}
&\bar{P} (u,t) = \frac{1}{i \Omega_R} \partial_t \bar{S}(u,t)  \nonumber \\
&= \frac{i}{\Omega_R}e^{- \frac{A}{2}t } \left ( \frac{A}{2} \cos (\Omega_R t)  + \Omega_R \sin(\Omega_R t)\right )  \bar{S}(u,t=\tau_d), \label{Psoln}
\end{align}
where we then see that with a sufficiently fast $\pi$-pulse (such that $2\Omega_R \tau_R  = \pi$) obeying $\gamma_{eg} (1+d) \tau_R \ll 2$, one can convert $S$ to $P$ without loss, yielding
\begin{align}
\bar{P}(u, \tau_R + \tau_d) \approx  i \bar{S}(u,t=\tau_d) \label{perfecttransfer}.
\end{align}

 The emitted read photon field can then be obtained by solving the set of equations in  (\ref{HeisenbergLangevinRetrieval}) after the fast read control field has ended (see Appendix A2), giving
\begin{align}
 \mathcal{E}_r(0,t)   =& i \sqrt{\frac{\gamma_{eg} d  }{c L}}  e^{-\gamma_{eg} t}\int _{0} ^L \! J_0\Big[ 2 \sqrt{\frac{\gamma_{eg} d}{L}t (L- z_1'') }    \Big] \nonumber \\
&   P(L- z_1'', \tau_R+ \tau_d) \, dz_1''. 
 \label{pipulseemissionforward}
\end{align}

Along with Eqn (\ref{perfecttransfer}) and noting that $\int_0^\infty e^{-\alpha x} J_\nu(2\beta \sqrt{x})J_\nu(2\gamma \sqrt{x}) dx = \frac{1}{\alpha} I_\nu(\frac{2\beta \gamma}{\alpha}) \text{exp}(-\frac{\beta^2 + \gamma^2}{\alpha})$ \cite{Gradshteyn07}, this emitted field then yields a retrieval efficiency given by Eqn (\ref{CompleteRetrieval}).

\subsection{Comparison}

We have seen that the proposed retrieval protocol yields a dependence on the spin shape, as described by Eqn (\ref{CompleteRetrieval}). Hence we now compare the retrieval efficiencies attainable with our protocol and compare them to the optimal ones. 

We can estimate the achievable efficiency of our protocol by choosing a write pulse duration  such that the resultant spin shape best fits the optimal spin shape. A good approximation to this write pulse duration is well described in \cite{Vivoli13}, and given by
\begin{align}
 \tau_W^{\text{approx}} = \frac{1}{\gamma_{eg}} \frac{1}{1 + \frac{d}{2}}.
\end{align}

\begin{center}
\begin{figure}[htbp!]
\includegraphics[width=0.97\linewidth]{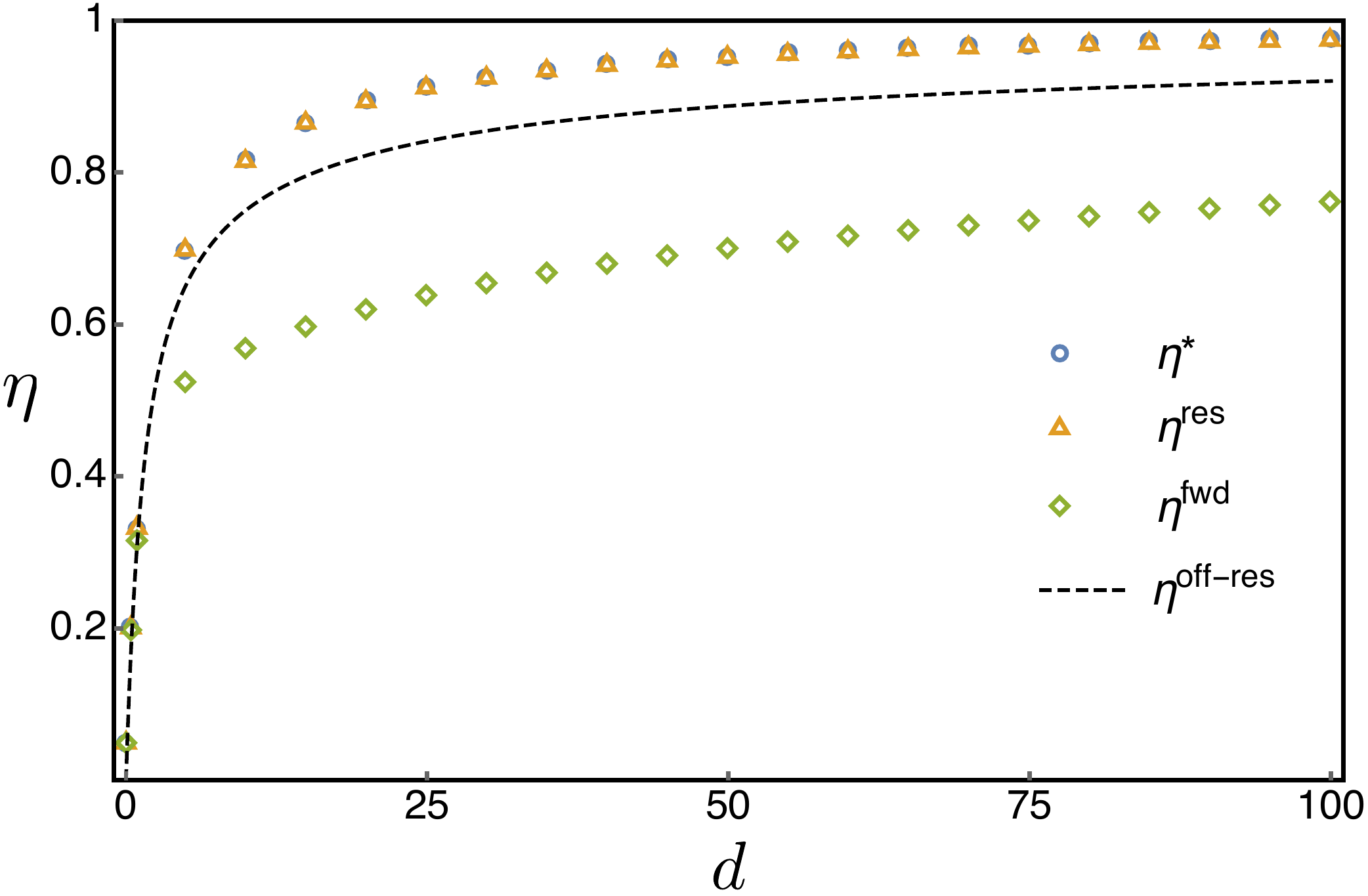}
\caption{Retrieval efficiency as a function of the optical depth.  Blue circles indicate the retrieval efficiency from the optimal spin wave. Yellow triangles (green diamonds) indicate the efficiency from backward (forward) retrieval for the proposed recipe that uses an exponentially rising write control field. The black dashed line indicates the retrieval efficiency using the standard approach with off-resonant write control fields.   }\label{0to100plot}
\end{figure}
\end{center}

We also compute the retrieval efficiencies $\eta^{\text{fwd}}$ that would be obtained if the resonant write pulse of duration $\tau_W^{\text{approx}}$ were to be followed by a co-propagating retrieve pulse instead. This would result in a situation where the spin wave would be far from optimal with respect to the retrieval direction.
In Fig \ref{0to100plot}, we compare the optimal efficiency $\eta^*$, the efficiencies $\eta^{\text{res}}$ and $\eta^{\text{fwd}}$ obtained with our proposal (from a spin wave created from a resonant exponential pulse with duration $\tau_W^{\text{approx}}$) together with the efficiency of the standard approach using far off-resonant write pulses, for which the efficiency is bounded by the complete retrieval efficiency from a flat spin wave \cite{GorshkovII}

\begin{align}
\eta^{\text{off-res}} = 1 - e^{-d}(I_0(d) + I_1(d)),
\end{align}
which we have verified numerically.
This retrieval efficiency is valid for retrieval from both the forward and backward directions from a flat spin wave.

Our proposal approaches optimal efficiencies, performing within $\sim10^{-3}$ of $\eta^*$ and compares favorably with respect to the standard off-resonant case. The improvement in efficiency is dependent on the optical depth, and we present some values in Table I.

\begin{table}[H]
\caption{Comparison of retrieval efficiency from different spin shapes \label{tabletable}}
\begin{ruledtabular}
\begin{tabular}{c c c c c}
$d$ & $\eta^{\text{fwd}} $&$\eta^{\text{off-res}}$&  $\eta^{\text{res}}$ & $\eta^{*}$ \\
    \hline \hline 
   0.1 & 0.0476 & 0.0476	& 0.0476	& 0.0476 \\
  1    & 0.3140 & 0.3263	& 0.3305 	& 0.3305 \\ 
  10   & 0.5671 & 0.7509	& 0.8134 	& 0.8142 \\
  20    & 0.6183 & 0.8227	& 0.8921 	& 0.8973 \\ 
  100  & 0.7600 & 0.9203	& 0.9728 	& 0.9745 \\  
\end{tabular}
\end{ruledtabular}

\end{table}

\subsection{Retrieval into a single mode}
For a single photon source to be useful, one needs to not only efficiently emit a single photon, but also to ensure that the given photon is emitted in a pure state. Here, we note that the number of emission modes $K$ can be estimated from an autocorrelation measurement, which gives  $g^{(2)} \sim 1+ \frac{1}{K}$ \cite{Sekatski12} (valid in the absence of detector noise and for small emission probabilities) for a process creating correlated photons in vacuum squeezed states.

 In the regime considered so far, we computed (see Appendix D for details)
\begin{align}
g^{(2)}(0) &= \frac{ \langle \mathcal{E}^\dagger_r(0,t) \mathcal{E}^\dagger_r(0,t) \mathcal{E}_r(0,t) \mathcal{E}_r(0,t) \rangle}{\langle \mathcal{E}^\dagger_r(0,t)  \mathcal{E}_r(0,t)  \rangle ^2 } \nonumber \\
&=2.
\end{align}
This is a good indication that the read photon field is emitted in a single mode, and hence that the conditional read field is a single photon in a pure state.

\section{Feasibility study of Rubidium-87} 
For a feasibility study we consider a $\Lambda$-system consisting of the following energy levels from the D-2 transition:  $|g \rangle  = | 5^2 \text{S}_{1/2}, \text{F}=2, m_\text{F}=2\rangle$, $|s \rangle = | 5^2 \text{S}_{1/2}, \text{F}=1, m_\text{F}=0 \rangle$ and $|e \rangle =| 5^2 \text{P}_{3/2} , \text{F}=2 , m_\text{F}=1 \rangle $. By taking into account the relevant branching ratios, we take $\gamma_{eg} = \frac{1}{12} (2 \pi )6.067 \text{ MHz} $ and $\gamma_{es} = \frac{1}{8} (2 \pi )6.067 \text{ MHz} $. For a sample with an optical depth $d=20$, a suitable write control field duration is given by $\gamma_{eg} \tau_W ^{\text{approx}} = 0.09 $. This implies a field duration of $\tau_W ^{\text{approx}} \sim$  29 ns. 

 Assuming that optical depth on the other transition is $\bar{d}=20$, and with a weak write control field such that $\Omega_W^{max} \tau_W =0.01$, within the short detection window $\tau_d\approx 0.1 \mu\text{s}$ the number of write photons is $n_w = 2\times10^{-4}$.
 
Subsequently, the retrieval pulse on the $|e\rangle$-$ |g\rangle$ transition requires a Rabi frequency of $\Omega_R  \gg  (2\pi) 5.3 \text{ MHz}   $, with a predicted retrieval efficiency of 89\%, essentially achieving $\eta^*$ (see Table \ref{tabletable}). This compares favorably to the retrieval efficiency from a flat spin wave $\eta^{\text{off-res}} =82\% $.

\section{Conclusion} 
In this work, we have discussed conditions for optimal generation of single photons from spontaneous Raman processes in cold atoms. We have proposed a detailed recipe to create single photons with efficiencies that compare favorably to standard strategies utilising flat spin waves. The recipe consists of first sending a resonant exponentially rising write control field onto an atomic sample of duration $ \tau_W \approx\gamma_{eg} ^{-1}(1+\frac{d}{2})^{-1} $. This heralds a spin wave that has a spatially varying form. Upon a fast $\pi$-retrieval, one obtains near-optimal retrieval efficiencies, and we find that the emitted single photons are in a pure state.  This proposal shows a convenient way to maximise the efficiency of single photon sources with given optical depths based on spontaneous Raman processes. This work could help in the implementation of the first quantum repeater protocol successfully outperforming the direct transmission of photons \cite{Sangouard11}.

\section*{ACKNOWLEDGMENTS}
We would like to acknowledge  Mikael Afzelius, Jean-Daniel Bancal, Lucas Beguin, Pau Farrera, Georg Heinze, Enky Oudot, Tan Peng Kian and Janik Wolters for useful discussions. Research at the University of Basel is supported by the Swiss National Science Foundation (SNSF) through the grant number PP00P2-150579 and the Army Research Laboratory Center for Distributed Quantum Information via the project SciNet. H. de R. aknowledges financial support by the  Spanish Ministry of Economy and Competitiveness (MINECO) and Fondo Europeo de Desarrollo Regional (FEDER) (FIS2015-69535-R), by MINECO Severo Ochoa through Grant No. SEV-2015-0522, by Fundaci\'o Cellex, and by CERCA programme/Generalitat de Catalunya.

\section*{APPENDIX A: RETRIEVAL PROCESS}
\subsection*{A1: Retrieval Emission Dynamics}
We begin from the Hamiltonian $H = H_0 + V$ (see \cite{GorshkovII}), where we consider an atomic sample of length L, and a classical field sent from the $z=L$ side of the sample. Choosing $|g\rangle $ to be the energy level reference for the atomic states, we have
\begin{align}
H_0 = \int d \omega \hbar \omega \hat{a}_\omega ^\dag \hat{a}_\omega  + \sum_{i=1} ^N (\hbar \omega_{s} \sigma^j _{ss} + \hbar \omega_{e} \sigma^j _{ee} )
\end{align}
\begin{align}
V =& -\hbar \sum _{i=1} ^N \Big( \Omega_R(t - \frac{L-z_i}{c}) \sigma^i_{es} e^{-i \omega_2 t}  e^{+i \omega_2(\frac{L-z_i}{c})} \nonumber  \\
 &+ g \sqrt{ \frac{L}{2 \pi c} }  \int d \omega \ a_\omega e^{i \omega \frac{L-z_i}{c}} \sigma^i_{eg} +H.c.) ,
\end{align}
\noindent where $\sigma_{\mu \nu}^i = | \mu \rangle _i \langle \nu | $ indicates atomic level operators for the $i$-th atom, and $a_w$ indicates the annihilation operator for the photonic mode at frequency $\omega$. $\omega_2 \  ( \omega_1)$ indicates the frequency of the read control (photon) field respectively.  Note that we are considering resonant pulses, so we have $\omega_1\  (\omega_2) = \omega_e\ (\omega_s)$.
Using 
\begin{align}
A&=\sum_{i=1}^N\Big[\hbar ( \omega_{1} - \omega_{2}) \sigma_{ss}^i + \hbar \omega_{1} \sigma_{ee}^i \Big] \nonumber \\
& \ \ \ +  \hbar \omega_1 \int \mathrm{d} \omega \  \mathcal{E}_r^\dag(z,t) \mathcal{E}_r(z,t) , \nonumber \\
U &= e^{-i At/\hbar} \nonumber ,
\end{align}
\noindent for the change of frame, then in the continuum limit, we obtain 
\begin{align}
H_{new} =& U^\dag H U - A \nonumber \\
=& \int \mathrm{d} \omega \  \hbar \omega a_\omega ^\dag a_\omega - \hbar \omega_1 \int \mathrm{d} z \  \mathcal{E}_r^\dag(z,t) \mathcal{E}_r(z,t)   \nonumber  \\
&+ \frac{N}{L} \int \mathrm{d} z \Big \{ -\hbar \Omega_R(z,t) \sigma_{es}(z,t) e^{+i \omega_2 \frac{L-z}{c}}+ h.c.   \nonumber \\
&\ \ \ \ \ \ \ \ \ \ \ \ \ \ \ \ - g \mathcal{E}_r (z,t) \sigma_{eg}(z,t) e^{+i \omega_1 \frac{L-z}{c}} + h.c. \Big \} , \nonumber 
\end{align}
\noindent where we have defined a real Rabi frequency $\Omega_R(z,t) = \Omega_R(t-\frac{L-z}{c})$ and $\mathcal{E}_r(z,t) = \sqrt{\frac{L}{2 \pi c}}  \int d \omega \ e^{i \omega_1 t}a_\omega e^{i (\omega-\omega_1) \frac{L-z}{c}}   $.
Using the field propagation equation along with the Heisenberg-Langevin equations of motion, we have in a moving coordinate frame, ignoring spinwave decoherence and the noise terms, and also considering that $\sigma_{gg}\sim 1$,
\begin{align}
\partial_z \mathcal{E}_r(z,t) &= -\frac{i g \sqrt{N}}{c} P(z,t) \nonumber \\
\partial_t P(z,t) &= -\gamma_{eg} P(z,t) + i g \sqrt{N} \mathcal{E}_r(z,t) + i \Omega_R(t) S(z,t) \nonumber \\
\partial_t S(z,t) &= i \Omega_R(t) P(z,t) , \label{retrieveHeisenbergLangevin}
\end{align}
\noindent where $g^2 N = \frac{d \gamma_{eg}c}{L}$,  $P(z,t) = \sqrt{N} \sigma_{ge}(z,t)e^{-i \omega_1 \frac{L-z}{c}}$ and $S(z,t) = \sqrt{N} \sigma_{gs}(z,t)e^{-i (\omega_1-\omega_2) \frac{L-z}{c}}$. 
In the continuum limit, the spin and field operators obey the following commutation relations
\begin{align}
&[\sigma_{\alpha \beta}(z,t) , \sigma_{\mu \nu}(z',t)] \nonumber \\
&= \frac{L}{N}\delta(z-z') ( \delta_{\beta \mu }\sigma_{\alpha \nu}(z,t) - \delta_{\nu \alpha}\sigma_{\mu \beta}(z,t)),\\
&[\mathcal{E}_{r}(z,t), \mathcal{E}_{r}^\dagger(z,t') ] = \frac{L}{c}\delta(t-t').
\end{align}
Rewriting Eqns (\ref{retrieveHeisenbergLangevin}) in the reverse direction e.g. $\bar{\mathcal{E}}_r(z',t) = \bar{\mathcal{E}}_r(L-z, t) = \mathcal{E}_r(z, t)$, and taking the Laplace transform from $z' \rightarrow u$ we obtain
\ba
\bar{\mathcal{E}}_r(u,t) &=& \frac{i g \sqrt{N}}{cu} \bar{P}(u,t) + \frac{1}{u} \bar{\mathcal{E}}_r (z'=0,t), \nonumber \\
\partial_t \bar{P}(u,t) &=& -\gamma_{eg} \bar{P}(u,t) + i g \sqrt{N} \bar{\mathcal{E}}_r (u,t) + i \Omega_R(t) \bar{S}(u,t),  \nonumber \\
\partial_t \bar{S}(u,t) &=& i \Omega_R(t) \bar{P}(u,t).
\ea
We can combine these three equations into a single differential equation, where we have ignored the boundary term $\bar{\mathcal{E}}_r(z'=0,t)$ since we send the read control field into the $z=L$ side of the atoms. On resonance ($\Delta$=0), let $A =\gamma_{eg} + \frac{g^2 N}{c u }  $ and $B = \Omega_R^2$ to see
\begin{align}
\ddot{\bar{S}}(u,t)  + A \dot{\bar{S}}(u,t) + B \bar{S}(u,t)=0 \label{spindiffeqn}.  
\end{align}
\subsection*{A2: Fast retrieval}
In the strong regime for the read control field, one requires $2|\Omega_R| \gg \gamma_{eg} (1+d)$, which implies
\begin{align}
2 \Omega_R  &\gg \gamma_{eg} (1 + \frac{d z'}{L }) \nonumber \\
\implies 2 \Omega_R  &\gg\gamma_{eg} (1 + \frac{d}{L u})  \nonumber   ,
\end{align}
which then yields the regime $4B \gg A^2$.

\noindent The solution for Eqn (\ref{spindiffeqn}) in this regime is 
\ba
\bar{S}(u,t) =  e^{-At/2} \cos(\Omega_R t) C_1(u) +  e^{-At/2}  \sin(\Omega_R t) C_2(u) \nonumber ,
\ea
\noindent where the initial condition implies
\ba
\bar{S}(u,t) =  e^{-At/2} \cos(\Omega_R t) \bar{S}(u, t=0) \nonumber .
\ea
One can then find the prepared polarization in terms of the intial spin condition,
\ba
&&\bar{P}(u,t) \nonumber \\
&=& \frac{1}{i \Omega_R} \partial_t \bar{S}(u,t) \nonumber  \\
&=& \frac{i}{\Omega_R}e^{- \frac{A}{2}t } \Big( \frac{A}{2} \cos (\Omega_R t)  + \Omega_R \sin(\Omega_R t) \Big)  \bar{S}(u,t=0) \nonumber .
\ea
In the limit where we have a sufficiently strong read control field $(2 \Omega_R \gg \frac{\pi}{2}\gamma_{eg}(1+d)  )$, the $\pi$-pulse is completed quickly and we obtain a lossless preparation of $\bar{P}(u,t)$ from $\bar{S}(u,t=0)$ in the form
\begin{align}
\bar{P}(u,\tau_R) &= i \bar{S}(u,t=0). \label{losslessP}
\end{align}
Once the polarization is prepared, we find the emission by solving for the dynamics in the absence of the laser,
\begin{align}
\partial_z \bar{\mathcal{E}}_r(z,t) &= -i \frac{g\sqrt{N}}{c}\bar{P}(z,t), \nonumber \\
(\partial_t + \gamma_{eg} ) \bar{P}(z,t) &= i g \sqrt{N} \bar{\mathcal{E}}_r(z,t).  \nonumber 
\end{align}
Taking the Laplace transform from $L-z = z' \rightarrow u$ and neglecting the boundary term since it does not contribute to the photon number, we have
\begin{align}
\bar{\mathcal{E}}_r(u,t) &= i \frac{g\sqrt{N}}{c u } \bar{P}(u,t), \nonumber \\
(\partial_t  + \gamma_{eg} )\bar{P}(u,t) &= i g \sqrt{N} \bar{\mathcal{E}}_r(u,t) =- \frac{g^2 N}{c u }\bar{P}(u,t) \nonumber .
\end{align}
This yields the evolution of $P(u,t)$ after its preparation from $S(u,t)$,
\begin{align}
\bar{P}(u,t) = e^{-(\gamma_{eg} + \frac{g^2 N }{c u})(t-\tau_R)}\bar{P}(u, \tau_R),
\end{align}
\noindent and gives an emitted field of 
\begin{align}
\bar{\mathcal{E}}_r(z',t) =& i \frac{g\sqrt{N}}{c} e^{- \gamma_{eg} (t -\tau_R)}  \nonumber \\
& \int_0 ^{z'} dz'' J_0 \Bigg[2\sqrt{\frac{g^2 N}{c} (t-\tau_R)(z' - z'')}\Bigg] \bar{P}(z', \tau_R) \nonumber  ,
\end{align}
where $J_n[x]$ refers the $n$-th Bessel function of the first kind. 
Now with $z' = L-z$, we require the field at $z=0$ for backward retrieval, and we finally obtain 
\begin{align}
\mathcal{E}_r(0,t) =& - \frac{g\sqrt{N}}{c} e^{- \gamma_{eg}t} \nonumber \\
& \int_0 ^{L} dz'' J_0 \Bigg[2\sqrt{\frac{g^2 N}{c} t(L - z'')}\Bigg] S(L-z'', 0)   \label{idealretrieve},
\end{align}
where we have used Eqn (\ref{losslessP}) for a lossless preparation.
\subsection*{A3: Slow retrieval}

In the weak regime for the read control field, one requires $2|\Omega_R| \ll \gamma_{eg} $, which implies 
\ba
2 \Omega_R &\ll& \gamma_{eg} (1 + \frac{d z}{L })  \nonumber \\
\implies 2 \Omega_R&\ll&\gamma_{eg} (1 + \frac{d}{L u})  \nonumber ,
\ea
which then yields the regime $4B \ll A^2$.

\noindent The solution for Eqn (\ref{spindiffeqn}) in this regime is  
\ba
\bar{S}(u,t) = e^{-\frac{1}{2}(A + \sqrt{A^2 -4B})t}C_u(1) + e^{-\frac{1}{2}(A - \sqrt{A^2 -4B})t}C_u(2). \nonumber 
\ea
When there is no laser ($B=0$), there should be no spinwave decay since we have considered zero spin wave decoherence, so we set $C_u(1)=0$ and obtain
\ba
\bar{S}(u,t) =  e^{-\frac{1}{2}(A - \sqrt{A^2 -4B})t}\bar{S}(u, t=0). \nonumber 
\ea

\noindent Now, in this regime when the Rabi frequency is small, we have  
\ba
e^{ - \frac{1}{2}(A - \sqrt{A^2 - 4B})t} &=&e^{ - \frac{1}{2}(A - A\sqrt{1-\frac{4B}{A^2}})t }  \nonumber \\
&\approx& e^{ - \frac{B}{A}t}  \nonumber \\
&=& e^{ -  \frac{\Omega^2}{\gamma_{eg} (1 + \frac{d}{L u})}t } \nonumber
\ea
This gives
\ba
\bar{S}(u,t) &=& e^{-Kt \frac{1}{1+ s/u } } \bar{S}(u, t=0), \nonumber 
\ea
where $K = \frac{\Omega_R^2}{\gamma_{eg}}$ and $s =\frac{d}{L}$.
One can proceed to find $\bar{P}(u,t) = \frac{1}{i \Omega_R}\partial_t \bar{S}(u,t)$ and $\bar{\mathcal{E}}(u,t) = i\frac{ g \sqrt{N}}{cu} \bar{P}(u,t)$, giving
\ba
\bar{\mathcal{E}}(u,t) = -\frac{g\sqrt{N}}{c} \frac{K}{\Omega_R}\Big[ \frac{1}{u + s}e^{-Kt + Kt(\frac{s}{s+u})}\Big]\bar{S}(u, t=0). \nonumber 
\ea
This yields
\begin{align}
\bar{\mathcal{E}}(z',t) &= - \frac{g\sqrt{N}}{c} \frac{K}{\Omega} e^{-Kt}  \nonumber \\
& \int_0^{z'} e^{-s(z'-z'')} I_0(2\sqrt{K t s(z'-z'')}) \bar{S}(z'', t=0) dz''  
\end{align}

One can then compute the retrieval efficiency from a single spin wave, and this is found to yield the optimal retrieval efficiency.
\begin{align}
&\int_0 ^\infty  dt \  \frac{c}{L} \langle \mathcal{E}^\dagger(0,t) \mathcal{E}(0,t)  \rangle  \nonumber \\
=& \int_0 ^\infty  dt \ \frac{d}{L^2} \frac{\Omega_R^2}{\gamma_{eg}} e^{-2Kt} \int_0 ^L dz_1''\  \int_0 ^L dz_2'' \  e^{-\frac{d}{L}(2L-z_1''-z_2'')}\nonumber \\
& I_0\Big(2\sqrt{K t \frac{d}{L}(L-z_1''}\Big) I_0\Big(2\sqrt{K t \frac{d}{L}(L-z_2''}\Big)   \nonumber  \\
&\langle S^\dagger(L-z_1'', t=0) S(L-z_2'', t=0) \rangle  \nonumber \\
=& \frac{1}{L} \int_0 ^L dz_1'' \ \frac{1}{L} \int_0 ^L dz_2'' \ \frac{d}{2} e^{-\frac{d}{2}\frac{(L-z_1'') + (L-z_2'')}{L}} \nonumber \\
&I_0\Big(d\sqrt{\frac{L-z_1''}{L}}\sqrt{\frac{L-z_2''}{L}}\Big)  \nonumber  \\
& \langle S^\dagger(L-z_1'', t=0) S(L-z_2'', t=0) \rangle  \nonumber 
\end{align}
where $I_n[x]$ denotes the $n$-th modified Bessel function of the first kind. We have made use of the fact that $I_n(z) = i^{-n} J_n(i z)$ and also $\int_0^\infty e^{-\alpha x} J_\nu(2\beta \sqrt{x})J_\nu(2\gamma \sqrt{x}) dx = \frac{1}{\alpha} I_\nu(\frac{2\beta \gamma}{\alpha}) \text{exp}(-\frac{\beta^2 + \gamma^2}{\alpha})$.

\section*{APPENDIX B: WRITE PROCESS}
\subsection*{B1: Heisenberg-Langevin equations for the atomic coherences}

The goal here is to first derive the expressions for the evolution of the atomic coherences in the write process.
\noindent We begin from the Hamiltonian $\bar{H} =\bar{H}_0 + \bar{V}$
\begin{align}
\bar{H}_0 = \int d \omega \hbar  \omega a_\omega ^\dag a_\omega  + \sum_{i=1} ^N (\hbar \omega_{s} \sigma^j _{ss} + \hbar \omega_{e} \sigma^j _{ee} )
\end{align}
\begin{align}
\bar{V} =& -\hbar \sum _{i=1} ^N \Big( \Omega_W(t - z_i/c) \sigma^i_{eg} e^{-i \omega_1(t-z_i/c)} \nonumber  \\
 &+ \bar{g} \sqrt{ \frac{L}{2 \pi c} }  \int d \omega \ \hat{a}_\omega e^{i \omega z_i /c} \sigma^i_{es} +H.c.) 
\end{align}

\noindent Using 
\begin{align}
\bar{A}=&\sum_{i=1}^N(\hbar \omega_{s} \sigma_{ss}^i + \hbar \omega_{e} \sigma_{ee}^i) \nonumber \\
 \ \ \ &+ \hbar \omega_2 \int \mathrm{d} z \  \mathcal{E}_w^\dag(z,t) \mathcal{E}_w(z,t), \nonumber \\
\bar{U} =& e^{-i \bar{A}t/\hbar}  \nonumber
\end{align}
for the change of frame, then in the continuum limit, we obtain
\begin{align}
\bar{H}_{new} =& \bar{U} ^\dagger \bar{H}\bar{U} - \bar{A}\\
=& \int \mathrm{d} \omega \  \hbar {a}_\omega ^\dag {a}_\omega - \hbar \omega_2 \int \mathrm{d} z \  \mathcal{E}_w^\dag(z,t) \mathcal{E}_w(z,t)   \nonumber  \\
&+ \frac{N}{L} \int \mathrm{d} z \big \{ -\hbar \Omega_W(t - z/c) \sigma_{eg}(z,t)e^{i \omega_1 z/c}  + h.c.\nonumber  \\
&\ \ \ \ \ \ \ \ \ \ \ \ \ \ \ \ - g \mathcal{E}_w (z,t) \sigma_{es}(z,t)e^{i \omega_2 z/c} + h.c. \Big \} , \nonumber 
\end{align}
\noindent where we have defined ${\mathcal{E}}_w(z,t) = \sqrt{\frac{L}{2 \pi c}} e^{i \omega_2 t}  \int d \omega \ {a}_\omega  e^{i (\omega-\omega_2) z /c}    $.

\noindent Assuming a real Rabi frequency $\Omega_W$, this yields the Heisenberg-Langevin equations as follows:
\begin{align}
\partial_t \sigma_{se} =& - \gamma_{es} \sigma_{se} + i \Omega_W e^{i \omega_1 z/c} \sigma_{sg} \nonumber \\
& -i \bar{g}\mathcal{E}_w e^{i \omega_2 z/c}(\sigma_{ee} - \sigma_{ss}) + F_{se} \nonumber \\
\partial_t \sigma_{sg} =& - \gamma_0 \sigma_{sg}+ i \Omega_W e^{-i \omega_1 z/c} \sigma_{se}  \nonumber \\
&- i \bar{g} \mathcal{E}_w e^{i \omega_2 z/c} \sigma_{eg} + F_{sg} \nonumber \\
\partial_t \sigma_{eg} =& - \gamma_{eg} \sigma_{eg} - i \Omega_W e^{-i \omega_1 z/c} \sigma_{gg} + F_{eg}, \label{writeequations}
\end{align}
\noindent where $\omega_1$ ($\omega_2$) indicates the frequency of the write control field (write photon field) respectively and $\bar{g}^2 N = \frac{\bar{d} \gamma_{es} c}{L}$. 

\subsection*{B2: Creating atomic coherences}
During the write process we account for possible depletion of the write laser intensity, and hence do not assume $\Omega_W(r,t)$ to be constant throughout the sample. As a result of the laser we create coherences between the $|g\rangle$-$|e\rangle$ transition, which forms the initial state for the write photon field. Here we proceed to find the atomic coherences prepared as a result of our exponential shaped resonant write control field.

\noindent For a sufficiently short write control field, the dynamics of the field and the atoms can be described with the dynamics along the $|g\rangle$-$|e\rangle$ transition. Ignoring the noise terms on $\sigma_{ge}$ and making the analogy between the classical and quantum fields on the $|g\rangle$-$|e\rangle$ transition, 
\ba
c \partial_z \Omega_W(z,t) &=& i g^2N \sigma_{ge}(z,t)  e^{-i \omega_1 z/c}, \nonumber  \\
\partial_t \sigma_{ge} &= & - \gamma_{eg}  \sigma_{ge} + i \Omega_W(z,t) e^{+i \omega_1 z/c} \sigma_{gg} \nonumber \\
&\approx& - \gamma_{eg}  \sigma_{ge} + i \Omega_W(z,t) e^{+i \omega_1 z/c} ,
\ea
\noindent where we have assumed that almost all atoms remain in the $|g\rangle$ level. 

\noindent Let us first assume a write control field with Rabi frequency $\Omega_W$ that begins at t=0. Taking the Laplace transforms from $t \rightarrow w$, we find
\begin{align}
\partial_z \Omega_W(z,w) &= i \frac{g^2N}{c} \sigma_{ge}(z,w)  e^{-i \omega_1 z/c},  \nonumber \\
\sigma_{ge}(z,w) &= \frac{1}{w + \gamma_{eg}} [i \Omega_W(z,w) e^{i \omega_1 z/c}   +\sigma_{ge} (z, t=0)]   \label{prepatomiccoh}.  
\end{align}
\noindent Insert the second equation of (\ref{prepatomiccoh}) into the first, and use the initial condition $\sigma_{ge}(z, t=0)=0$ to obtain
\ba
\partial_z \Omega_W (z,w) = - \frac{g^2 N}{c} (\frac{1}{w + \gamma_{eg}}) \Omega_W(z,w) , \nonumber 
\ea
\noindent yielding
\ba
\Omega_W(z,w) = e^{- \frac{g^2N}{c} (\frac{1}{w + \gamma_{eg}})z} \Omega_W (z=0,w). \nonumber 
\ea
\noindent Insert this into the second equation of (\ref{prepatomiccoh}) to obtain 
\ba
\sigma_{ge} (z,w) = (i e^{i \omega_1 z/c}) \Big[ \frac{1}{w + \gamma_{eg}} e^{- \frac{\bar{g}^2N }{c} \frac{1}{w + \gamma_{eg}}z} \Omega_W(z=0,w)  \Big ] . \nonumber 
\ea

\noindent After inverting the Laplace transform, we now shift the limits to consider a write control field with support on negative times, giving
\begin{align}
\sigma_{ge}(z,t)
=& (i e^{i \omega_1 z/c}) \int_{-\infty} ^t e^{- \gamma_{eg}(t-t_1'')}  \nonumber \\
&J_0\Big[2\sqrt{\frac{\gamma_{eg} d}{L}(t-t_1'') z} \Big]  \Omega_W(z=0, t_1'') dt_1'',
\end{align}
\noindent where $J_n(x)$ indicates the $n$-th Bessel function of the first kind.

\noindent Thus, with an exponential write control field $\Omega_W (0,t) = \Omega_W ^{max} e^{t/\tau_W} $ sent up to $t$=0, we evaluate the atomic coherence at $t$=0 with the help of $\int_0 ^\infty e^{- A t} J_0[2\sqrt{Bt}] dt = \frac{1}{A}e^{-B/A}$ and finally obtain 
\ba
\sigma_{ge} (z,0) = e^{i\omega_1 z/c} \theta_0 e^{-\frac{\alpha z}{2}} \label{prepared},
\ea

\noindent where $\theta_0 = i \frac{\Omega_{max} \tau_W}{1 + \gamma_{eg} \tau_W}$ and $\alpha /2 =  d \frac{\gamma_{eg} \tau_W}{1+ \gamma_{eg}\tau_W}\frac{1}{L}$.

\subsection*{B3: Write Photon Emission}
After the preparation of atomic coherences, we begin to see spontaneous emission from the $|e\rangle$ level. Along with the field propagation equation, the relevant Heisenberg-Langevin equations are
\ba
c\partial_{z} \mathcal{E}_w &=& i \bar{g} N e^{-i\omega_2 z/c} \sigma_{se}(z,t) ,\nonumber\\
\partial_{t} \hat{\sigma}_{es} &=& -\gamma_{es} {\sigma}_{se}  -i \bar{g} {\mathcal{E}}_w e^{i \omega_2 z/c} ({\sigma}_{ee} - {\sigma}_{ss}) + {F}_{se} . \nonumber 
\ea

\noindent Defining $Q^\dagger = \sqrt{N}e^{-i \omega_2 z/c} \sigma_{se}$, we will consider the write emission for short detection times. Using (\ref{prepared}) we thus replace ${\sigma}_{ee} - {\sigma}_{ss}$ with its mean value at position z and $t=0$ to obtain
\ba
c\partial_{z} \mathcal{E}_w(z, t) &=& i\bar{g}\sqrt{N} Q^\dagger (z,t) , \label{FieldEqn} \nonumber\\
\partial_{t} Q^\dagger(z, t) &=&  -\gamma_{es}Q^\dagger(z,t) -i\bar{g}\sqrt{N} |\theta_0|^2 e^{-\alpha z} \mathcal{E}_w(z,t) \nonumber  \\
&&+ F_Q^\dagger (z, t).  \label{QdagforLangevinForm}
\ea

\noindent Performing first the Laplace transform in space ($z \rightarrow s$)
\begin{align}
&s\mathcal{E}_w(s,t) - \mathcal{E}_w(z=0,t) = A \ Q^\dagger(s ,t) ,  \nonumber\\
&\partial_{t} Q^\dagger (s, t) = -\gamma_{es} Q^\dagger (s, t) + B\ \mathcal{E}_w(s + \alpha, t) \nonumber \\
&\ \ \ \ \ \ \ \ \ \ \ \ \ \ \ \ + F_Q^\dagger(s, t),  \nonumber  
\end{align}
and then in time ($t \rightarrow \omega$), we get
\begin{align}
&s\mathcal{E}_w(s,\omega) - \mathcal{E}_w(z=0,\omega) = A \ Q^\dagger(s ,\omega) ,\nonumber\\
&Q^\dagger (s, \omega)  = \frac{1}{\gamma_{es} + \omega}\{  B\ \mathcal{E}_w(s + \alpha, \omega)  \nonumber\\
&\ \ \ \ \ \ \ \ \ \ \ \ \ \ + F_Q^\dagger(s, \omega) + Q^\dagger(s, t=0)\} ,\nonumber   
\end{align}
where $A = i \frac{\bar{g}\sqrt{N}}{c}$ and $B = -i\bar{g}\sqrt{N} \theta_0^2$.

\noindent Substituting the second line into the first, we eliminate $Q(s,\omega)$ and are left with a boundary term in $Q$: 
\begin{align}
&s \mathcal{E}(s,\omega) - \mathcal{E}(z=0, \omega) = (\frac{A}{\gamma_{es} + \omega})[ B\ \mathcal{E}(s + \alpha, \omega) \nonumber \\
& + F_Q^\dagger(s, \omega) + Q^\dagger(s, t=0)]  . \nonumber 
\end{align}

\noindent The following formula also holds with a shift from $s$ to $s + \alpha$:
\begin{align}
&(s+ \alpha) \mathcal{E}_w(s+ \alpha,\omega) - \mathcal{E}_w(z=0, \omega)  \nonumber \\
=& (\frac{A}{\gamma_{es} + \omega})\Big[ B\ \mathcal{E}_w(s +2  \alpha, \omega)  + F_Q^\dagger(s+ \alpha, \omega)\nonumber \\
&+ Q^\dagger(s + \alpha, t=0) \Big]  . \nonumber 
\end{align}

\noindent By substituting $\mathcal{E}_w(s+ \alpha, \omega)$ into the previous equation we can find $\mathcal{E}(s, \omega)$ in terms of $\mathcal{E}_w(s+ 2\alpha , \omega)$, and by taking the substitution into the n-th step we have 
\begin{align}
\mathcal{E}_w(s, \omega) =& K(\omega)^n D(n) \mathcal{E}_w(s + n \alpha, \omega) \nonumber \\
&+ \frac{1}{B} \sum^n _{j=1} K(\omega)^j D(j) F_Q^\dagger \left(s + (j-1)\alpha, \omega \right)\nonumber  \\
&+ \frac{1}{B} \sum^n _{j=1} K(\omega)^j D(j) Q^\dagger \left(s + (j-1)\alpha, t'=0 \right) \nonumber \\
&+ \left[ K(\omega)  \right]^{-1} \sum^n _{j=1} \left [ K(\omega)^j D(j) \right] \mathcal{E}_w(z'=0, \omega),\nonumber
\end{align}
where $K(\omega) = \frac{AB}{\gamma_{es} + \omega}$ and $D(n) =  { \displaystyle \prod_{k=0}^{n-1} \frac{1}{s + k \alpha}}$.

\noindent Taking the limit of $n \rightarrow \infty$, the first term disappears, and we proceed to perform the inverse transform $s\rightarrow z$. With a shift in the index $j$, $\mathcal{L}^{-1}[D(j+1)] = \frac{1}{j!} \left( \frac{1-e^{-\alpha z}}{\alpha} \right)^j$ and the shifting property of the Laplace Transform, 
\begin{align}
&\mathcal{E}_w (z, \omega) \nonumber \\
 =& \frac{1}{B} \sum_{j=0}^\infty K(\omega)^{j+1}  \int_0 ^{z} \frac{1}{j!} \left( \frac{1-e^{-\alpha(z-z'')}}{\alpha} \right)^j \nonumber \\
 & e^{-j\alpha z''} F_Q^\dagger(z'', \omega) \  dz'' \nonumber \\ 
&+ \frac{1}{B} \sum_{j=0}^\infty K(\omega)^{j+1}  \int_0 ^{z} \frac{1}{j!} \left( \frac{1-e^{-\alpha(z-z'')}}{\alpha} \right)^j \nonumber \\
& e^{-j\alpha z''} Q^\dagger(z'', t=0) \  dz'' \nonumber \\ 
&+ \frac{1}{K(\omega)} \sum_{j=0}^\infty K(\omega)^{j+1} \frac{1}{j!} \left( \frac{1-e^{-\alpha(z)}}{\alpha} \right)^j \mathcal{E}_w(z'=0, \omega) \nonumber \\
 =& \frac{A}{\gamma_{es} + \omega} \int_0^{z} e^{[  \frac{1}{\gamma_{es} + \omega} M(z,z'') e^{-\alpha z''} ]} F_Q^\dagger (z'', \omega) dz''\nonumber  \\
&+\frac{A}{\gamma_{es} + \omega} \int_0^{z} e^{[  \frac{1}{\gamma_{es} + \omega} M(z,z'') e^{-\alpha z''} ]} Q^\dagger (z'', t'=0) dz''\nonumber  \\
&+ e^{\frac{1}{\gamma_{es} + \omega} M(z, 0)} \mathcal{E}_w (z=0, \omega), \nonumber 
\end{align}

\noindent where $M(z', z'' ) = \frac{AB}{\alpha} \left[ 1-e^{-\alpha(z'-z'')} \right]$.

\noindent Finally, noting that 
\begin{align} 
&\mathcal{L} ^{-1} \left[ \frac{1}{\gamma_{es} + \omega} e^{\frac{A}{\gamma_{es} + \omega}} \right]  \nonumber \\
=& e^{-\gamma_{es} t} \left[  \sqrt{ At } ^{(-1)} I_1(2\sqrt{At})  + I_2 (2\sqrt{At}) \right],\nonumber \\
&\mathcal{L} ^{-1} \left[ e^{\frac{A}{\gamma_{es} + \omega}} \right]  \nonumber \\
 =& e^{-\gamma_{es} t} \left[ \sqrt{\frac{A}{t}} I_1(2\sqrt{At}) + \delta(t) \right],\nonumber 
\end{align}
 
\noindent  we get 
\begin{align}
 \mathcal{E}_w (z, t)  =& A \int_0^{z} \int_0^{t} e^{-\gamma_{es} (t-t_1'')}  H_1[\alpha, z, z_1'', t, t_1'']  \nonumber  \\
 &F_Q^\dagger (z_1'',t_1'') dt_1'' dz_1'' \nonumber  \\
 &+ A \int_0^{z} e^{-\gamma_{es} (t)}  H_1[\alpha, z, z_1'', t, 0]  Q^\dagger (z_1'',0)  dz_1'' \nonumber   \\
 &+ \int_0^{t} e^{-\gamma_{es}(t - t'')}  H_2(\alpha, z,0, t, t'') \mathcal{E}_w(0, t'')  dt'' \nonumber \\
 &+ \mathcal{E}_w(0,t)
\end{align}
\noindent  where
\begin{align} 
&H_1 \left(\alpha, z_1,z_2,t_1,t_2 \right)    = I_0 \left[2\sqrt{M(z_1, z_2) e^{-\alpha z_2} (t_1 - t_2) } \right], \nonumber \\
&H_2 \left(\alpha, z_1,z_2,t_1,t_2 \right) = \sqrt{\frac{M(z_1,z_2)}{t_1 -t_2}} \nonumber \\
&  I_1 \left[2\sqrt{M(z_1, z_2) e^{-\alpha z_2} (t_1 - t_2) } \right]. \nonumber
  \end{align}

\subsection*{B4: Number of write photons}
Computing the photon flux requires the commutation relations for $Q$ and a 2-point noise correlation function involving $F_Q$. In a short time window $\tau_d$ where $\sigma_{ee} - \sigma_{ss} $ is not changing, and with the Einstein relations (see Chpt 15.5 of  \cite{MeystreOptics}), the Langevin equations for system operators can be written
\begin{align}
\dot{A}_\mu = D_\mu(t) + F_\mu(t). \label{LangevinForm}
\end{align} 

\noindent The corresponding memoryless noise correlations for operators $\mu$  and $\nu$ are such that
\begin{align}
\langle F_{\mu }(t') F_{\nu}(t'') \rangle = 2 \langle D_{\mu \nu} \rangle  \delta(t' - t''),
\end{align}
where
\begin{align}
2 \langle D_{\mu \nu} \rangle = - \langle A_\mu D_\nu \rangle - \langle D_\mu A_\nu \rangle +
 \frac{d}{dt} \langle A_\mu A_\nu \rangle  .
\end{align}

\noindent Thus, identifying terms in Eqn (\ref{QdagforLangevinForm}) with terms in Eqn (\ref{LangevinForm}), we make use of 
\begin{align}
[Q(z,t), Q^\dagger(z',t)] &= N[\sigma_{es}(z,t) , \sigma_{se}(z',t)]\nonumber \\
&=  L\delta(z-z') |\theta_0|^2 e^{-\alpha z'},
\end{align}
\noindent then we make use of the fact that $\langle Q^\dagger(z,t) Q(z',t) \rangle$ right after our preparation of atomic coherences is zero, giving
$ \langle Q(z,t) Q ^\dagger(z',t) \rangle = L\delta (z - z') |\theta_0|^2 e^{-\alpha z'}$.

\noindent Then one obtains
\begin{align}
2 \langle D_{Q, Q^\dagger} \rangle  &= 2\gamma_{es} L |\theta_0|^2 e^{-\alpha z} \delta(z-z') ,
\end{align}
\noindent yielding
\begin{align}
\langle F_Q(z,t)F_Q^\dagger(z', t') \rangle = 2\gamma_{es} L |\theta_0|^2 e^{-\alpha z} \delta(z-z') \delta(t-t'), 
\end{align}

\noindent valid when $\sigma_{ee} - \sigma_{ss}$ is not changing.

\noindent This yields a photon flux of
\begin{align}
&\frac{c}{L}\langle \mathcal{E}_w^\dagger (L,t) \mathcal{E}_w(L,t) \rangle\nonumber \\
 =& \frac{c}{L}\frac{\bar{g}^2N}{c^2}  \int_0^{L} e^{-2\gamma_{es} t } H_1[\alpha, L, z_1'',t, 0]^2 V|\theta_0|^2 e^{-\alpha z_1''} dz_1'' \nonumber \\
&+\frac{c}{L}\frac{\bar{g}^2N}{c^2} \int_0^{t} \int_0^{L} e^{-2\gamma_{es} (t - t_1'')} H_1[\alpha, L, z_1'',t, t_1'']^2 \nonumber \\
&\ \ \ 2 \gamma_{es} L |\theta_0|^2 e^{-\alpha z_1''} dz_1'' dt_1''  \nonumber
\end{align}
 
\noindent  For sufficiently short detection times $
 \tau_d \ll \frac{1}{2 \gamma_{es}} $, the noise contribution (second term) can be ignored, and furthermore when the photon number is much smaller than 1 $(\tau_d \ll \{\frac{\bar{g}^2N}{c}|\theta_0|^2 (\frac{1-e^{-\alpha L}}{\alpha})\}^{-1})$ we can consider just the leading term in the series expansion, and observe a constant flux. 
\begin{align}
&\frac{c}{L}\langle \mathcal{E}_w^\dagger (L,\tau_d) \mathcal{E}_w(L,\tau_d) \rangle \nonumber  \\
=& \frac{\bar{g}^2 N}{c} |\theta_0|^2 \int_0^{L}  \left(I_0 \left[ 2 \sqrt{M[ L, z_1'']e^{-\alpha z_1''}\tau_d} \right] \right)^2 e^{-\alpha z_1'' } dz_1'' \nonumber \\
\approx & \frac{\bar{g}^2 N}{c} |\theta_0|^2 \int_0^{L}   \left( 1 + 2 M[ L, z_1'']e^{-\alpha z_1''}\tau_d  + O(\tau_d^2) \right) \nonumber \\
& \  e^{-\alpha z_1'' } dz_1'' \nonumber \\
=& \frac{\bar{g}^2 N}{c} |\theta_0|^2 \frac{1 - e^{-\alpha L}}{\alpha} . 
\end{align} 

\noindent We therefore obtain a photon number of $\frac{\bar{g}^2 N}{c} |\theta_0|^2 \frac{1 - e^{-\alpha L}}{\alpha} \tau_d$ within this short detection window.  This is precisely the excited atom fraction multiplied by $\bar{d} \gamma_{es} \tau_d$, since the fraction of atoms that were excited after the write pulse is $\frac{1}{L} \int_0 ^L \langle  \sigma_{ee}(z,0) \rangle dz = \frac{1}{L} \int_0 ^L |\theta_0|^2 e^{-\alpha z} dz  = |\theta_0|^2 \frac{1 - e^{-\alpha L}}{\alpha L}$.

 \subsection*{B5: Number of prepared spins}

 We start with the description of the spin operator from Eqn (\ref{writeequations}), by defining $S = \sqrt{N}\sigma_{gs} e^{-i(\omega_1 - \omega_2)z/c}$ and replacing $\sigma_{eg}(z,t) $ by its mean value $\theta_0^* e^{-\alpha z/2} e^{-i\omega_1 z/c}$ 
\begin{align}
&(\partial_{t} + \gamma_0)\hat{S}^\dagger(z,t) - F_S^\dagger(z,t) \nonumber  \\
=&-i\bar{g}\sqrt{N} e^{ik_w r} \mathcal{E}_w(z,t)\sigma_{eg} \nonumber \\
=& -i \bar{g}\sqrt{N} \mathcal{E}_w(z,t)  \left[ \theta_0^* e^{-\alpha z/2}\right].  \nonumber \label{SpinEqn}
\end{align}
Take the Laplace transform from $t \rightarrow \omega$ to see
 \begin{align}
&(\omega+ \gamma_0)S^\dagger(z, \omega) - S^\dagger (z, t=0) -F_S^\dagger(z,\omega) \\
&=C (z) \mathcal{E}_w(z, \omega) , \nonumber 
\end{align}
 where $C(z) = -i\bar{g}\sqrt{N} \theta_0^* e^{-\alpha z/2}$.
Then we have
\ba
S^\dagger(z, \omega) &=& \frac{C(z)}{\omega+ \gamma_0}  \mathcal{E}_w(z, \omega) \nonumber  \\ 
&+& \frac{1}{\omega + \gamma_0} S^\dagger(z, t=0) \nonumber \\
&+& \frac{1}{\omega + \gamma_0} F_S^\dagger(z, \omega) ,
\ea
\noindent and noting that $\mathcal{L}^{-1}[\frac{1}{\omega + \gamma_0}] = e^{-\gamma_0 t}$ yields
\ba
S^\dagger(z, t) &=& C(z)\int_0^{t} e^{-\gamma_0(t - t')} \mathcal{E}_w(z, t') dt' \nonumber \\
&+& e^{-\gamma_0 t} S^\dagger(z, t=0) \nonumber \\
&+& \int_0^{t} e^{-\gamma_0 (t - t')} F_S^\dagger (z,t') dt' ,
\ea
\noindent where the field expression $\mathcal{E}$ from the previous subsection is required.
Ignoring terms that do not show up in the normal ordered $\langle S^\dagger S \rangle$, we have 
\ba
S^\dagger(z,t) &=& C(z)\int_0^{t}dt' e^{-\gamma_0(t - t')}  \nonumber \\
&&  \Bigg{\{} \int_0 ^{t'} dt'' e^{-\gamma_{es} (t' - t'')}  H_2(\alpha, z,0,t',t'') \mathcal{E}_w(0,t'') \nonumber \\
&&\ \ \  + \mathcal{E}_w(0,t') \Bigg{\}} \label{spinafterwrite}.
\ea
Computing $\langle S^\dagger S \rangle$ requires the commutator $[\mathcal{E}_w(z,t), \mathcal{E}_w^\dagger(z',t')] = L \delta[z - z' - c(t-t')]$ and yields 4 terms. In the short time window where one can ignore the atomic dephasing ($\tau_d \ll \frac{1}{2 \gamma_0}, \frac{1}{2 \gamma_{es}} $), and also where the photon number is much smaller than 1  $(\tau_d \ll \{\frac{\bar{g}^2N}{c}|\theta_0|^2 (\frac{1-e^{-\alpha L}}{\alpha})\}^{-1})$, only one term dominates (the term independent of $H_2$).
The number of spins is then equivalent to the photon number
\ba
\frac{1}{L} \int_0 ^L \langle S^\dagger (z,
\tau_d) S (z,\tau_d) \rangle dz \approx \frac{\bar{g}^2N }{c} |\theta_0|^2 \frac{1-e^{-\alpha L}}{\alpha} \tau_d. \nonumber 
\ea

\section*{APPENDIX C: PHASE MATCHING}
By assuming the retrieval process to perform retrieval from the exact same spin wave function $S(z,t)$ that has been created by the write pulse, we have assumed the degeneracy of the two metastable states $|g\rangle$ and $|s\rangle$. In general, the metastable states could have different energies which would lead to a read process from $S(z,t)e^{2i (\omega_e - \omega_s) z/c}$. However, this effect is negligible in the regime $|\omega_e - \omega_s|  \frac{L}{c} \ll 1 $.

\section*{APPENDIX D: SECOND ORDER COHERENCE}
In order to ascertain if the read photon field is of a single mode, we compute the unconditional autocorrelation function of the read photon field at time 0 is
\begin{align}
g^{2}(0) &= \frac{ \langle \mathcal{E}^\dagger_r(0,t) \mathcal{E}^\dagger_r(0,t) \mathcal{E}_r(0,t) \mathcal{E}_r(0,t) \rangle}{\langle \mathcal{E}^\dagger_r(0,t)  \mathcal{E}_r(0,t)  \rangle ^2 } \nonumber .
\end{align} 

\noindent Assuming that the read and write photon fields are described by a two mode squeezed state, the $g^{(2)} (0)$ is of the form $1 + \frac{1}{K}$, where K indicates the number of modes \cite{Sekatski12}.

\noindent In the regime we consider, where we have a short detection time and a fast readout, developing the numerator of the $g^{(2)} $ function leads to the term
\begin{align}
& \langle \mathcal{E}_w(0,t_a) \mathcal{E}_w(0,t_b) \mathcal{E}^\dagger_w(0,t_c) \mathcal{E}^\dagger_w(0,t_d) \rangle  \nonumber \\
 = & \langle \mathcal{E}_w(0,t_a) \Big[ \mathcal{E}^\dagger_w(0,t_c) \mathcal{E}_w(0,t_b) + \frac{L}{c}\delta(t_b - t_c) \Big] \mathcal{E}_w(0,t_d)  \rangle   \nonumber \\
 = & \Big(\frac{L}{c}\Big)^2 \delta(t_a - t_c) \delta(t_b - t_d)  + \Big(\frac{L}{c}\Big)^2 \delta(t_a - t_d) \delta(t_b - t_c) , \nonumber 
\end{align} 
\noindent which yields 
\begin{align}
g^{2}(0) &= \frac{ 2 \langle \mathcal{E}^\dagger_r(0,t)  \mathcal{E}_r(0,t)  \rangle ^2}{\langle \mathcal{E}^\dagger_r(0,t)  \mathcal{E}_r(0,t)  \rangle ^2 } =2 \nonumber 
\end{align} 
\noindent where we have used Eqn (\ref{idealretrieve}) and the leading term of Eqn (\ref{spinafterwrite}).

\begin{appendix}

\end{appendix}

\end{document}